\documentclass[longauth]{aa}

\usepackage{graphicx}
\usepackage{graphics}

\usepackage{longtable}
\usepackage{txfonts}

\usepackage{epsf,epsfig}
\usepackage[dvips]{color}
\usepackage{color}
\usepackage{amssymb,amsfonts}
\usepackage{latexsym}

\usepackage{tabularx}
\usepackage{multirow}

\usepackage{natbib}
\bibpunct{(}{)}{;}{a}{}{,} 

\begin{document}

\newcommand{\vla}{VLA}
\newcommand{\most}{MOST}
\newcommand{\park}{Parkes}
\newcommand{\atc}{ATCA}

\newcommand{\rosat}{{\it ROSAT}}
\newcommand{\rxt}{{\it RXTE}}
\newcommand{\xmm}{{\it XMM-Newton}}
\newcommand{\chan}{{\it Chandra}}
\newcommand{\sax}{{\it BeppoSAX}}
\newcommand{\asca}{{\it ASCA}}
\newcommand{\suz}{{\it Suzaku}}

\newcommand{\intg}{{\it INTEGRAL}}
\newcommand{\ib}{IBIS}
\newcommand{\spi}{SPI}
\newcommand{\isg}{ISGRI}
\newcommand{\ibisg}{IBIS/ISGRI}

\newcommand{\swi}{{\it SWIFT}}
\newcommand{\bat}{BAT}

\newcommand{\gro}{{\it CGRO}}
\newcommand{\cpt}{COMPTEL}
\newcommand{\egr}{EGRET}

\newcommand{\agile}{{\it AGILE}}
\newcommand{\fermi}{{\it Fermi}}
\newcommand{\hess}{H.E.S.S.}



\newcommand{\un}[1]{~\hspace{-1pt}\ensuremath{\mathrm{#1}}}

\newcommand{\am}{$^{\prime}$}
\newcommand{\as}{$^{\prime\prime}$}
\newcommand{\gammaray}{$\gamma$-ray}
\newcommand{\gammarays}{$\gamma$-rays}
\newcommand{\xray}{X-ray}
\newcommand{\xrays}{X-rays}


\newcommand{\dieter}[1]{{\color{blue}  #1}}
\newcommand{\helene}[1]{{\color{green} #1}}


\newcommand{\psr}{PSR~J1357$-$6429}
\newcommand{\snc}{G309.8$-$2.6}
\newcommand{\hessj}{HESS~J1356$-$645}
\newcommand{\rxj}{RX~J1713.7$-$3946}
\newcommand{\velajr}{RX~J0852$-$4622}
\newcommand{\rcw}{RCW~86}



\def\edot{$\dot{{\rm E}}$}

\def\ks{km s$^{-1}$}
\def\kms{$\mathrm {km s}^{-1}$}
\def\d{$^\circ$}
\def\m{$^\prime$}
\def\s{$^{\prime\prime}$}
\def\hh{$^{\mathrm h}$}
\def\mm{$^{\mathrm m}$}
\def\second{$^{\mathrm s}$}
\def\cm3{cm$^{-3}$}
\def\pp{$^{\prime\prime}$}
\def\msun{M$_\odot$}

\def\eg{{\it e.g.~}}
\def\etal{et~al.~}
\def\ie{{\em i.e.~}}


\title{Discovery of the source HESS~J1356-645 associated 
       with the young and energetic PSR~J1357-6429}


\tiny{
\author{HESS Collaboration
\and A.~Abramowski \inst{1}
\and F.~Acero \inst{2}
\and F.~Aharonian \inst{3,4,5}
\and A.G.~Akhperjanian \inst{6,5}
\and G.~Anton \inst{7}
\and A.~Balzer \inst{7}
\and A.~Barnacka \inst{8,9}
\and U.~Barres~de~Almeida \inst{10}\thanks{supported by CAPES Foundation, Ministry of Education of Brazil}
\and Y.~Becherini \inst{11,12}
\and J.~Becker \inst{13}
% ...
}
}


\offprints{mrenaud@lupm.univ-montp2.fr, facero@in2p3.fr}

\institute{
Universit\"at Hamburg, Institut f\"ur Experimentalphysik, Luruper Chaussee 149, D 22761 Hamburg, Germany \and
Laboratoire Univers et Particules de Montpellier, Universit\'e Montpellier 2, CNRS/IN2P3,  CC 72, Place Eug\`ene Bataillon, F-34095 Montpellier Cedex 5, France \and
Max-Planck-Institut f\"ur Kernphysik, P.O. Box 103980, D 69029 Heidelberg, Germany \and
Dublin Institute for Advanced Studies, 31 Fitzwilliam Place, Dublin 2, Ireland \and
National Academy of Sciences of the Republic of Armenia, Yerevan  \and
Yerevan Physics Institute, 2 Alikhanian Brothers St., 375036 Yerevan, Armenia \and
Universit\"at Erlangen-N\"urnberg, Physikalisches Institut, Erwin-Rommel-Str. 1, D 91058 Erlangen, Germany \and
Nicolaus Copernicus Astronomical Center, ul. Bartycka 18, 00-716 Warsaw, Poland \and
CEA Saclay, DSM/IRFU, F-91191 Gif-Sur-Yvette Cedex, France \and
University of Durham, Department of Physics, South Road, Durham DH1 3LE, U.K. \and
Astroparticule et Cosmologie (APC), CNRS, Universit\'{e} Paris 7 Denis Diderot, 10, rue Alice Domon et L\'{e}onie Duquet, F-75205 Paris Cedex 13, France \thanks{(UMR 7164: CNRS, Universit\'e Paris VII, CEA, Observatoire de Paris)} \and
Laboratoire Leprince-Ringuet, Ecole Polytechnique, CNRS/IN2P3, F-91128 Palaiseau, France \and
Institut f\"ur Theoretische Physik, Lehrstuhl IV: Weltraum und Astrophysik, Ruhr-Universit\"at Bochum, D 44780 Bochum, Germany
% ...
}


\normalsize

\date{Received ; accepted}

\abstract{Several newly discovered very-high-energy (VHE; E $>$ 100\un{GeV}) $\gamma$-ray sources in 
the Galaxy are thought to be associated with energetic pulsars. Among them, middle-aged ($\gtrsim$ 
10$^{4}$ yr) systems exhibit large centre-filled VHE nebulae, offset from the pulsar position, which result 
from the complex relationship between the pulsar wind and the surrounding medium, and reflect the past evolution 
of the pulsar.}{Imaging Atmospheric Cherenkov Telescopes (IACTs) have been successful in revealing extended emission 
from these sources in the VHE regime. Together with radio and \xray~observations, this observational window 
allows one to probe the energetics and magnetic field inside these large-scale nebulae.}{\hess, with its large field 
of view, angular resolution of $\lesssim$ 0.1\d~and unprecedented sensitivity, has been used to discover a large population of 
such VHE sources. In this paper, the \hess~data from the continuation of the Galactic Plane Survey (-80\d~$<$ $\ell$ 
$<$ 60\d, $|b|$ $<$ 3\d), together with the existing multi-wavelength observations, are used.}{A new VHE $\gamma$-ray 
source was discovered at R.A.~(J2000) = 13\hh56\mm00\second, Dec.~(J2000) = $-$64\d30\m00\s~with a 2\m~statistical error 
in each coordinate, namely \hessj. The source is extended, with an intrinsic Gaussian width of (0.20 $\pm$ 0.02)\d. 
Its integrated energy flux between 1 and 10\un{TeV} of 8 $\times$ 10$^{-12}$ erg cm$^{-2}$ s$^{-1}$ represents 
$\sim$ 11\% of the Crab Nebula flux in the same energy band. The energy spectrum between 1 and 
20\un{TeV} is well described by a power law dN/dE $\propto$ E$^{-\Gamma}$ with photon index $\Gamma$ = 2.2 
$\pm$ 0.2$_{stat}$ $\pm$ 0.2$_{sys}$. The inspection of archival radio images at three frequencies and the analysis
of \xray~data from \rosat/PSPC and \xmm/MOS reveal the presence of faint non-thermal diffuse emission coincident with \hessj.}
{\hessj~is most likely associated with the young and energetic pulsar \psr~(d = 2.4\un{kpc}, $\tau_c$ = 7.3\un{kyr} 
and \edot~= 3.1 $\times$ 10$^{36}$ erg s$^{-1}$), located at a projected distance of $\sim$ 5 pc from the centroid 
of the VHE emission. \hessj~and its radio and \xray~counterparts would thus represent the nebula resulting from 
the past history of the \psr~wind. In a simple one-zone model, constraints on the magnetic field strength in the nebula 
are obtained from the flux of the faint and extended \xray~emission detected with \rosat~and \xmm. \fermi-LAT upper limits 
in the high-energy (HE; 0.1--100\un{GeV}) domain are also used to constrain the parent electron spectrum. From the low 
magnetic field value inferred from this approach ($\sim$ 3--4 $\mu$G), \hessj~is thought to share many similarities with 
other known $\gamma$-ray emitting nebulae, such as Vela~X, as it exhibits a large-scale nebula seen in radio, \xrays~and 
VHE gamma-rays.}


\titlerunning{Discovery of \hessj}

\maketitle


\section{Introduction}
\label{s:intro}

The survey of the Galactic Plane conducted with the \hess~(High Energy Stereoscopic System) experiment since 2004, 
covering essentially the whole inner Galaxy \citep{c:survey1,c:survey2}, has led to the discovery of about fifty new 
sources in the very-high-energy (VHE; E $>$ 100\un{GeV}) $\gamma$-ray domain \citep{c:hinton09}. 
A significant fraction of these sources remain without any clear counterpart at lower (radio and \xrays) energies \citep{c:darks}. 
Pulsar wind nebulae (PWNe), which represent the largest population of identified Galactic VHE sources so far, could 
naturally account for a large fraction of these unclassified sources \citep{c:dejager09}. These bubbles of relativistic 
particles and magnetic field are created when the ultra-relativistic pulsar wind interacts with the surrounding medium 
\citep[see][for recent reviews]{c:gaensler06,c:bucciantini08}. Such interaction leads to the formation of a so-called 
termination shock, which is thought to be the site of particle acceleration beyond hundreds of TeV. Luminous 
nebulae are thus observed across the entire electromagnetic spectrum, in the synchrotron emission from radio to hard 
\xrays, and through inverse Compton processes and potentially $\pi^{0}$ decay from p-p interactions \citep{c:amato03,c:bb03,c:horns06}, 
in the VHE domain. Radio and \xray~observations have revealed the complex morphology of the inner PWN structure at 
the arcsecond scale \citep{c:gaensler06}. Moreover, \hess~has proven itself to be capable of measuring, in HESS~J1825-137 
\citep{c:j1825}, spatially resolved spectra at the tens of arcmin scale. These VHE observations, together with the on-going 
\fermi-LAT (Large Area Telescope) observations in the high-energy (HE; 0.1--100\un{GeV}) gamma-ray domain \citep{c:fermi09}, 
permit one to probe the particle spectra in these sources, and provide a unique approach to determining the average 
magnetic field strength \citep[see][for a discussion in this regard]{c:dd09}.

Two classes of VHE PWNe have recently emerged, based on their observational properties: {\it young} systems, such as the 
Crab nebula \citep{c:crab}, G0.9+0.1 \citep{c:g09}, MSH~15$-$52 \citep{c:msh1552} and the newly discovered \hess~sources 
associated with the Crab-like pulsars of G21.5$-$0.9 and Kes~75 \citep{c:youngs}, and the {\it middle-aged} sources 
(\ie with characteristic ages $\tau_{c}$ $\gtrsim$ 10$^{4}$ yr\footnote{Bearing in mind that the characteristic age 
may not reflect the true pulsar age, as this is valid only when the braking index n = 3 and when the pulsar's initial 
spin period is much smaller than that observed today.}), as exemplified by Vela~X \citep{c:velax}, HESS~J1825$-$137 
\citep{c:j1825}, HESS~J1718$-$385 and HESS~J1809$-$193 \citep{c:j1809}. In the former case, the VHE emission is usually 
unresolved and centered on the pulsar, while in the latter case, these VHE PWNe were found to be significantly extended and
offset from the pulsar position. The large differences in the measured sizes of the VHE and \xray~emission regions, ranging from 
a factor of a few to $\sim$ 100, can be explained by the difference in the cooling timescales of particles radiating in 
the two domains \citep[see {\it e.g.}][]{c:j1825}. The evolution of the Supernova Remnant (hereafter, SNR) blastwave into an 
inhomogeneous interstellar medium \citep[hereafter, ISM;][]{c:blondin01} and$/$or the high velocity of the pulsar \citep{c:vds04} 
may both explain these large offset centre-filled VHE sources as being the ancient nebulae from the history of the pulsar 
wind inside its host SNR \citep{c:dd09}.

This paper deals with the results from the \hess~observations and data analysis of one of these VHE PWN candidates, 
\hessj. This source lies close to \psr, a young ($\tau_{c}$ = 7.3 kyr) and energetic (with a spin-down luminosity 
\edot~= 3.1 $\times$ 10$^{36}$ erg s$^{-1}$) 166\un{ms} pulsar, discovered during the Parkes multibeam survey of the 
Galactic Plane \citep{c:camilo04}. At a distance of 2.4 $d_{2.4}$ kpc (with $d_{2.4}$ = d/2.4\un{kpc}), as inferred 
from dispersion measure, \psr~may be, after the Crab, the nearest young ($\tau_{c}$ $<$ 10$^{4}$ yr) pulsar known. 
Follow-up \xray~observations with \xmm~and \chan~have been performed \citep{c:esposito07,c:zavlin07}. Using radio 
ephemerides, an indication of a high \xray~pulsed fraction (p$_f$ $\gtrsim$ 50\%) was found by \citet{c:zavlin07}. 
The \xray~spectrum of \psr~is well described by a non-thermal (magnetospheric, $\Gamma$ = 1.4 $\pm$ 0.5) component 
plus a blackbody radiation at 0.16\un{keV} \citep{c:esposito07}, the latter being detected in only a few young 
radio pulsars. The fraction of the spin-down energy channeled into the observed 0.5-10\un{keV} flux of 
$\sim$ 7.4 $\times$ 10$^{-5}$ $d^{2}_{2.4}$ represents one of the lowest ever observed amongst the 
rotation-powered pulsars. As noticed by \citet{c:esposito07}, \psr~shares similarities in this regard 
with the Vela pulsar and PSR~B1706$-$44, for which thermal emission has also been detected. These authors 
found a 3 $\sigma$ marginal evidence of diffuse emissive region at the 10\s~scale in the \xmm/EPIC 2-4\un{keV} 
band and set a 3 $\sigma$ 2-10\un{keV} upper limit on a putative PWN with \chan~data of 2.8 $\times$ 
10$^{31}$ $d^{2}_{2.4}$ erg s$^{-1}$. On the other hand, \citet{c:zavlin07} reported the presence 
of a faint tail-like emission extending northeast at a distance of $\sim$ 2\s~from \psr, with a 
0.5-10\un{keV} luminosity of 2.3 $\times$ 10$^{31}$ $d^{2}_{2.4}$ erg s$^{-1}$. Since the proper motion 
of the pulsar has not yet been measured, the nature of this elongated feature is still under debate. 

The \hess~observations, data analysis and the characteristics of \hessj~are provided in section \ref{s:obsres}. 
Results from the analysis of archival radio continuum and \xray~(\rosat/PSPC and \xmm/MOS) data are presented 
in section \ref{s:counterparts}, together with those recently obtained from \fermi-LAT observations and
presented in the companion paper of \citet{c:lemoine11}. A general discussion is given in section \ref{s:discu}.

\section{\hess~observations and results}
\label{s:obsres}

\hess~comprises four identical 12\un{m} diameter IACTs located in the Khomas Highland of Namibia at a 
height of 1800\un{m} above sea level. Sensitive to $\gamma$-rays above $\sim$ 100\un{GeV}, the 
\hess~array commonly achieves an event-by-event angular resolution of $\lesssim$ 0.1\d~and a relative energy 
resolution of $\sim$ 15\%. The observations discussed here were first taken between April and June 2005, 
as part of the on-going \hess~Galactic Plane survey which now covers the band -80\d~$<$ $\ell$ $<$ 60\d~in longitude 
and $|b|$ $<$ 3\d~in latitude. Additionally, two sets of dedicated observations were taken in May 2006 and 
March 2007 in the direction of the pulsar \psr~in so-called {\it wobble mode} \citep{c:daum97}, where data 
are taken with an alternating offset from the target position of typically $\pm$ 0.7\d. Events within 
3\d~of \psr~and recorded when at least three of the four telescopes were operational were considered. After 
applying quality selection to discard data affected by unstable weather conditions or problems related to the hardware, 
the resulting dataset was analyzed using the standard \hess~survey analysis scheme \citep{c:survey1}. A radius 
of the on-source region $\theta_{cut}$ of 0.22\d, a ring background region with a mean radius of 0.8\d~and the 
{\it hard cuts} were used. These cuts include a minimum requirement of 200 photo electrons per shower image, 
parametrized using the Hillas moment-analysis technique \citep{c:hillas96}, for $\gamma$-ray selection. The 
zenith angles of the observations range from 39\d~to 54\d~(with a mean value of 45\d) inferring a mean energy 
threshold of 800\un{GeV}. The acceptance-corrected live time is $\sim$ 10 h at the position of the VHE emission.

\begin{figure}[!htb]
\centering

  \includegraphics[width=0.48\textwidth]{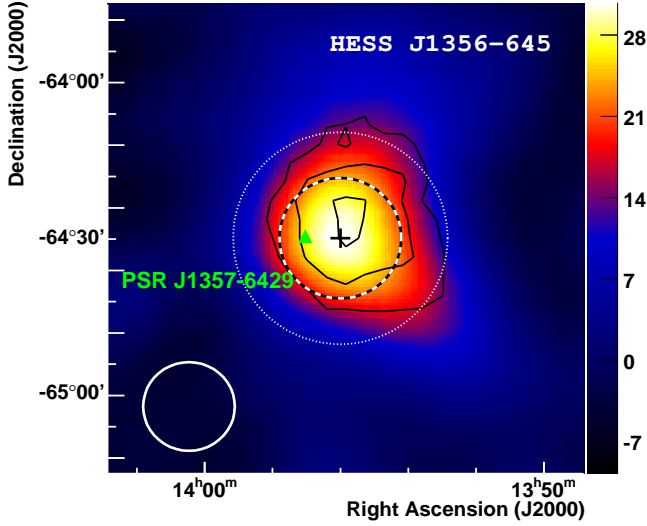}

  \caption{Image of the VHE $\gamma$-ray excess centered on \hessj, smoothed with a Gaussian of width 0.13\d. 
           The linear colour scale is in units of excess counts per arcmin$^{2}$. 
           The black contours correspond to the 5, 7 and 9 $\sigma$ levels for an integration radius 
	   of 0.22\d. The black cross indicates the best fit position of the source centroid together with 
           its statistical error. The intrinsic Gaussian source size is shown by the white dashed circle. 
           The dotted circle indicates the region of spectral extraction (see Figure \ref{f:ima2}). 
           The position of \psr~is marked with a green triangle. The white solid circle represents the 
	   68~\%~containment radius of the resulting point spread function.}
  \label{f:ima1}

\end{figure}

\vspace{-0.0cm}

The \hess~excess count image, smoothed with a Gaussian profile of width 0.13\d~in order to mitigate statistical 
fluctuations, is shown in Figure \ref{f:ima1}. An excess of VHE emission is found with a pre-trials peak significance 
of 10 $\sigma$, \ie 8.5 $\sigma$ after taking into account the number of trials \citep[see {\it e.g.}][]{c:survey1}. 
The centroid position and intrinsic Gaussian width of the source were determined by fitting the uncorrelated excess 
map with a symmetric Gaussian convolved by the nominal \hess~point spread function (PSF) for this dataset (with a 
68\% containment radius of $\sim$ 0.08\d). This led to a best-fit position of R.A. (J2000) = 13\hh56\mm00\second, 
Dec. (J2000) = $-$64\d30\m00\s~with statistical error of 2\m~for each coordinate, and an intrinsic width of 0.20\d 
$\pm$ 0.02\d$_{stat}$, as shown by the black cross and the dashed circle in Figure \ref{f:ima1}, respectively. 
Therefore, \hessj~is clearly an extended source, well described by a Gaussian profile at the level of available 
statistics.

\begin{figure}[!htb]

  \includegraphics[width=0.48\textwidth]{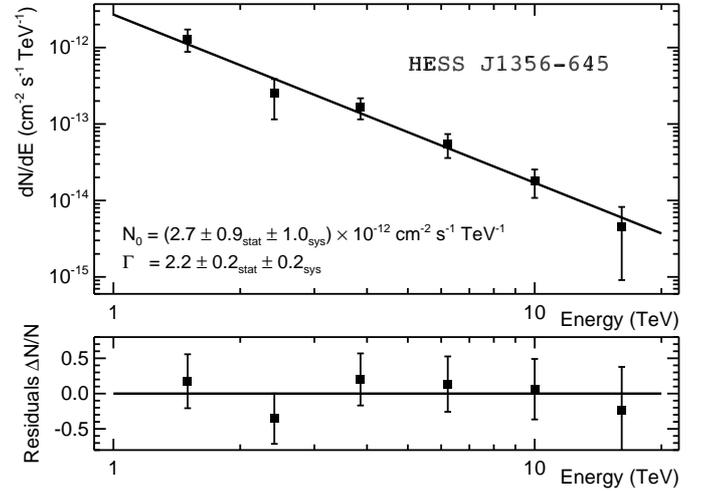}

  \caption{Differential energy spectrum of \hessj, extracted
           within the dotted circular region shown in Figure \ref{f:ima1}.
           Events with energies between 0.8 and 30\un{TeV} were
           binned in order to reach a significance of at least 2
           $\sigma$ per resulting bin. The data points were fitted
           with a power law whose best fit is shown with the black
           solid line. Also shown are the residuals in the bottom
           panel.}
  \label{f:ima2}

\end{figure}

The spectral analysis was performed on observations at less than 2\d~from the best-fit position of 
\hessj, to avoid any systematical effect arising from larger uncertainties in the reconstructed energy
for showers which are reconstructed far off-axis. The source spectrum was determined within a circular 
region of 0.34\d~radius, as shown by the dotted circle in Figure \ref{f:ima1}. This region represents an 
$\sim$80\% source enclosure, which is a compromise between the optimal signal to noise ratio and the 
independence to the source morphology. Using the same cuts as for the imaging analysis, the background 
is estimated using the {\it reflected-region} technique, where background events are selected from 
circular off-source regions of same angular size and offset from the observation position as the on-source 
region \citep{c:berge07}. The resulting spectrum of \hessj~shown in Figure \ref{f:ima2} is well described 
between 1 and 20\un{TeV} by a power-law of the form dN/dE = N$_{0}$ (E/1\un{TeV})$^{-\Gamma}$ with photon 
index $\Gamma$ = 2.2 $\pm$ 0.2$_{stat}$ $\pm$ 0.2$_{sys}$ and flux normalization at 1\un{TeV} N$_{0}$ = 
(2.7 $\pm$ 0.9$_{stat}$ $\pm$ 0.4$_{sys}$) $\times$ 10$^{-12}$ cm$^{-2}$ s$^{-1}$ TeV$^{-1}$ ($\chi^{2}$/dof 
= 1.8/4, with an associated p-value of 0.77). Adding an exponential cutoff to the power-law does not improve 
the fit. A lower limit on the cutoff energy of 3.5\un{TeV} at the 95\% confidence level was derived. The 
1--10\un{TeV} integrated energy flux of 8 $\times$ 10$^{-12}$ erg cm$^{-2}$ s$^{-1}$ represents $\sim$ 11\% 
of the Crab Nebula flux \citep{c:crab}, in the same energy band.

\vspace{-0.5cm}

\section{Search for counterparts}
\label{s:counterparts}

\begin{center}
\begin{figure*}[!htb]
\centering

  \includegraphics[width=0.96\textwidth]{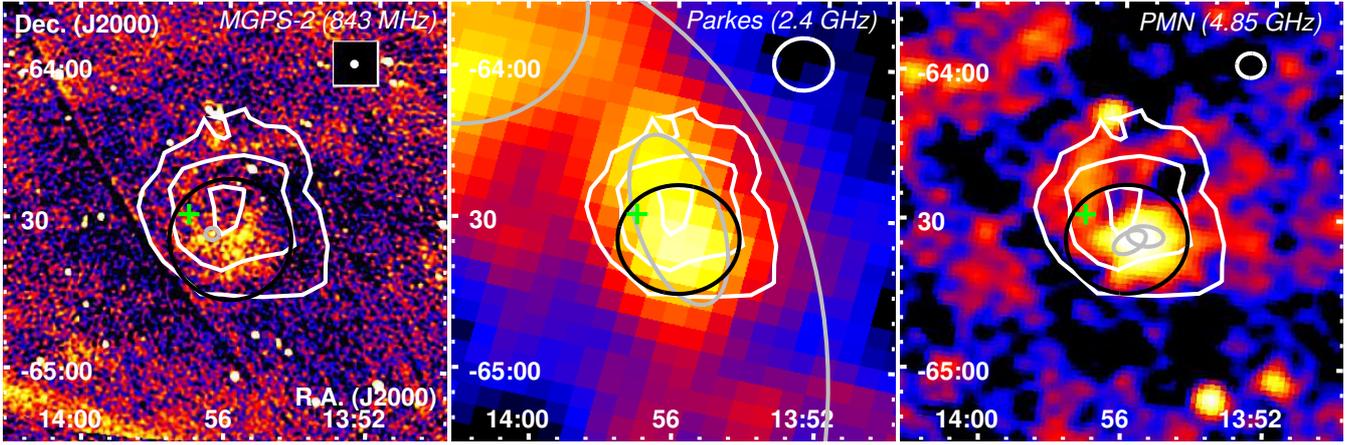}

  \caption{Radio images (in square root scale) centered on \hessj. The colour scale 
           is such that the blue-to-red transition occurs at about the 
           respective measured rms (see Table \ref{t:radio}). In each case, 
	   the beamsize of the instrument is depicted by the white circle in 
	   the upper right corner. The position of \psr~is marked with a green 
	   cross and the white contours represent the levels of the \hess~correlated
           significance, as shown in Figure \ref{f:ima1}. The black circle
           denotes the region within which flux densities were calculated.
           {\it Left:} Molonglo Observatory Synthesis Telescope
           (MOST) image at 843\un{MHz} from the MGPS-2 \citep{c:murphy07}. 
	   The grey circle denotes the position of the catalogued point-like 
	   source MGPS~J135623-643323. {\it Middle:} Parkes image at 
	   2.4\un{GHz} \citep{c:duncan95}. The three grey ellipses mark the 
	   positions of the SNR candidates G309.8-2.6 (in the centre), G310.6-2.0 
	   and G310.5-3.5 (only partly shown in the image) found by \citet{c:duncan97}. 
	   {\it Right:} Parkes-MIT-NRAO (PMN) image at 4.85\un{GHz} \citep{c:griffith93}. 
           The grey ellipses mark the positions of the two catalogued extended sources, 
	   PMN~J1355-6433 and PMN~J1355-6435 \citep{c:wright94}.}
  \label{f:ima3}

\end{figure*}
\end{center}

\vspace{-0.5cm}

\subsection{General considerations}
\label{ss:counterparts_gnrl}

Catalogues of Galactic SNRs \citep{c:green09}, HII regions \citep{c:paladini03}, star-forming complexes 
\citep{c:russeil03} and WR stars \citep{c:vdh01} do not list any potential VHE emitters in the vicinity of 
\hessj. However, at a projected distance of $\sim$ 7\m~(\ie~5 $d_{2.4}$ pc) from the centroid position of \hessj, 
\psr~appears to be the most plausible counterpart, within the context of VHE PWN candidates described in section 
\ref{s:intro}, given that its spin-down luminosity \edot~= 3.1 $\times$ 10$^{36}$ erg s$^{-1}$ would imply a
conversion efficiency of L$_{{\rm 1-10 TeV}}$/\edot~= (0.9-2.8) $\times$ 10$^{-3}$ $d^{2}_{2.4}$. Moreover,
\citet{c:camilo04} have shown that \psr~is located near the radio SNR {\it candidate} \snc~(see section 
\ref{ss:counterparts_radio}), previously discovered and catalogued as an elongated area of enhanced radio 
emission by \citet{c:duncan97}, from the \park~2.4\un{GHz} survey of the Southern Galactic Plane 
\citep[][see section \ref{ss:counterparts_radio}]{c:duncan95}.

The distance estimate of \psr~would result in a VHE intrinsic source size of $\sim$ 9 $d_{2.4}$ \un{pc}. The 
observed offset between \psr~and the centroid of the VHE emission could be accounted for as a result of either a high proper 
motion of the pulsar \citep{c:vds04}, or a density gradient in the surrounding medium \citep{c:blondin01}. In the former case, 
the required transverse velocity is $\sim$ 650 $d_{2.4}$ $\tau^{-1}_{7.3}$ km s$^{-1}$, where $\tau_{7.3}$ is the true 
age of the pulsar relative to its characteristic age of 7.3 kyr. Such a velocity is relatively large in comparison with 
the mean value reported by \citet{c:hobbs05} (with a mean 2D speed for pulsars with $\tau_{c} <$ 3\un{Myr} of 307 $\pm$ 
47 km s$^{-1}$) from a sample of 233 pulsar proper motions, but similar to what has been measured for a few PSRs. In the 
latter case, the location of \psr, $\sim$ 105 $d_{2.4}$ pc from the Galactic Plane, \ie~beyond the thin disc known to be 
highly inhomogeneous, would argue against the scenario of \citet{c:blondin01}, where the offset PWN would be explained by 
the asymmetry in the SNR geometry due to the inhomogeneities in the ISM. However, the lack of available molecular line 
data in the direction of \psr~prevents one from drawing firm conclusions in this regard.

Regardless of the physical explanation for the observed offset between \psr~and the VHE centroid, the system should have 
already experienced the reverse shock interaction phase, where the reverse shock has collided with the entire shock surface 
bounding the pulsar wind. After this phase, an offset PWN from the initial energetic stage of the pulsar wind is formed. 
\hessj~would then correspond to such a nebula radiating in the VHE domain. According to the hydrodynamical model of \citet{c:vds04}, 
this phase occurs on a timescale of $\lesssim$ 10$^{3}$ E$^{-1/2}_{51}$ (M$_{ej}$/M$_{\odot}$)$^{5/6}$ n$^{-1/3}_{0}$ yrs, 
where E$_{51}$ is the total mechanical energy of the SN explosion in units of 10$^{51}$ erg, M$_{ej}$ is the mass 
in the SN ejecta, and n$_{0}$ is the ambient number density in units of cm$^{-3}$. Even in the case of a 
core-collapse event (M$_{ej}$ = 10 M$_{\odot}$, with E$_{51}$ and n$_{0}$ equal to unity), this timescale is 
smaller than the characteristic age of \psr, which renders this scenario plausible.

\subsection{Radio continuum observations}
\label{ss:counterparts_radio}

As discussed in section \ref{s:intro}, PWNe are commonly detected in \xrays~and radio. Interestingly enough, 
the above-mentioned SNR candidate \snc~turns out to be located at only $\sim$ 0.1\d~away from \hessj. Archival 
radio images from the Molonglo Galactic Plane Survey (MGPS-2) at 843\un{MHz} \citep{c:murphy07}, from the 
Parkes 2.4\un{GHz} Survey \citep{c:duncan95} and from the Parkes-MIT-NRAO (PMN) survey at 4.85\un{GHz} 
\citep{c:griffith93} have been inspected in order to shed light on the nature of this SNR candidate. The main 
characteristics of these three surveys are reported in Table \ref{t:radio}.

\begin{table}[!htb]

\caption{Radio continuum observations towards \hessj. For each dataset, the centre frequency, the angular
resolution, the noise measured over a region of 1\d~in size centered on the VHE source and the nominal noise 
(in parentheses) are provided. Flux densities and errors have been measured within the black circle shown 
in Figure \ref{f:ima3}. Note that the flux density measured at 2.4\un{GHz} with Parkes has been subtracted 
from the underlying emission of G310.5-3.5 (see text).} \label{t:radio}

\centering
\begin{tabular}{|c|c|c|c|}
\hline

 & MGPS-2 & Parkes & PMN \\
\hline

Frequency (GHz)            &      0.843      &       2.4       &       4.85      \\
rms (mJy beam$^{-1}$)      &     1.8 (1.6)   &     90 (17)     &       14 (7)    \\
PSF (FWHM, arcmin)         &      0.8        &      10.4       &       4.9       \\
Flux density \& error (Jy) & 0.53 $\pm$ 0.04 &  1.5 $\pm$ 0.5  & 0.54 $\pm$ 0.05 \\

\hline
\end{tabular}

\end{table}

Images from these three radio surveys and centered on \hessj~are shown in Figure \ref{f:ima3}, with the 
position of \psr~denoted by the green cross and the \hess~significance contours from Figure \ref{f:ima1} 
overlaid in white. Extended emission is seen in each image, partly coincident with \hessj. On the left 
image, the grey circle marks the position of the point-like source MGPS~J135623-643323, with a flux density 
of 18.2 $\pm$ 2.8\un{mJy}, as measured by \citet{c:murphy07}. \citet{c:duncan97} have catalogued a list of 
SNR candidates from the Parkes 2.4\un{GHz} Survey. The selection was based on the source morphology, 
on the absence (or faint level) of coincident thermal emission by inspecting the IRAS images at 60 
$\mu$m, and on the level of polarization. Three of them, namely \snc~(at the centre of the image, 
15\m~$\times$ 35\m~in size), G310.6-2.0 (on the north-western region, 45\m~in diameter) and G310.5-3.5 
(2.7\d~$\times$ 3.5\d~in diameter), with respective flux densities of 3.9 $\pm$ 1.0, 11 $\pm$ 3 and 
19 $\pm$ 5\un{Jy}, are shown in grey in Figure \ref{f:ima3} (middle). Based on the 
polarized intensity images, \citet{c:duncan97} suggested that the first two SNR candidates may be 
unrelated to the larger G310.5-3.5, but there is no distance estimate for any of these three sources. 
On the right image, the two grey ellipses denote the positions of the two extended sources 
PMN~J1355-6433 and PMN~J1355-6435, with respective flux densities of 158 $\pm$ 11 and 140 $\pm$ 10\un{mJy}, 
as catalogued by \citet{c:wright94}. The black circle shown in Figure \ref{f:ima3} represents the region of 
spectral extraction (with a radius of 0.18\d) and was chosen to enclose most of the diffuse radio emission 
seen in the MGPS-2 and PMN images. It should be noted that the MOST telescope does not detect structures 
on angular scales larger than 20-30 arcmin \citep{c:murphy07}, and that such large sources are also partially 
suppressed during the PMN data reduction \citep{c:condon93}. This would explain why the larger SNR candidates 
G310.6-2.0 and G310.5-3.5 are not detected in the MGPS-2 and PMN images, while present in the image processed 
by \citet{c:duncan95} at 2.4\un{GHz}. 

In the MGPS-2 image\footnote{\texttt{\tiny http://www.astrop.physics.usyd.edu.au/mosaics/}}, the flux density 
of the point-like source MGPS~J135623-643323 was removed, after summing all the pixels inside the black circle 
and correcting for the beam. With an average brightness ($\sim$ 3 mJy beam$^{-1}$) of only two times the local
rms noise, the flux density of this extended emission at 843\un{MHz} (see Table \ref{t:radio}) must 
be considered with caution.

In the Parkes 2.4\un{GHz} image\footnote{\texttt{\tiny http://www.atnf.csiro.au/research/surveys/2.4Gh\_Southern/}}, 
\snc~lies above the broad and faint shell-like emission of the very large SNR candidate G310.5-3.5, which might be 
responsible for much of the diffuse emission, as mentioned by \citet{c:duncan97}. Therefore its flux density as well as 
its morphology have to be taken into consideration. Assuming that the three SNR candidates are unrelated to each other, 
the underlying contribution of G310.5-3.5 to the final flux density within the region of interest was estimated by 
modelling G310.5-3.5 as a uniform elliptical shell of 19\un{Jy}, with external sizes of 2.7\d~$\times$~3.5\d~as given 
by \citet{c:duncan97} and internal sizes half as large. This underlying emission amounts to $\sim$ 0.48 $\pm$ 0.13\un{Jy} 
and was subtracted from the total flux density (corrected for the beam). 

In the PMN image at 4.85\un{GHz}\footnote{\texttt{\tiny ftp://ftp.atnf.csiro.au/pub/data/pmn/surveys/}}, the two 
catalogued sources have not been considered as such in the calculation, given that some complex diffuse objects like the
one shown in Figure \ref{f:ima3} (right) may have been considered as multiple sources by the automatical procedure, 
which was optimized to primarily detect point sources \citep{c:wright94}. Instead, individual fluxes have been 
simply summed within the black circle and the flux density was corrected for the beam. 

The flux densities and respective errors at these three frequencies are reported in Table \ref{t:radio}. Fitting these 
data points with a power-law gives a normalization at 1\un{GHz} of 0.54 $\pm$ 0.04\un{Jy} and a slope $\alpha$ = 0.01 
$\pm$ 0.07 (where S$_{\nu}$ $\propto$ $\nu^{\alpha}$), with $\chi^{2}$/dof = 1.14/1. As seen in Table \ref{t:radio}, 
the 2.4\un{GHz} flux density is $\sim$ 2 $\sigma$ above the baseline formed by the two data points at 0.843 and 
4.85\un{GHz}. However, as presented above, estimating the contribution of the large and faint shell-like 
emission from G310.5-3.5 is a difficult task, and the flux density measured at 2.4\un{GHz} has certainly 
to be taken as an upper limit. For instance, the tail-like structure north-east to the central bright 
emission, originally considered by \citet{c:duncan97} as part of \snc, could simply be a region of
enhanced shell emission within G310.5-3.5. In any case, the 2.4\un{GHz} measurement only influences the 
normalization of the radio spectrum. A power-law fit on the two other data points, with a normalization 
at 1\un{GHz} fixed to 0.54\un{Jy}, gives a consistent spectral index of $\alpha$ = 0.01 $\pm$ 0.06.
Changing the size of the region defined for calculating the flux densities does not change the results
significantly. Such a flat radio spectrum suggests \snc~to be a PWN rather than a shell-type SNR given that
typical indices for PWNe are -0.3 $\lesssim$ $\alpha$ $\lesssim$ 0 \citep{c:gaensler06}, while those for shell-type
SNRs are usually close to -0.5 \citep{c:green09}. The lack of evidence for a shell-type morphology in the MGPS-2 
and PMN images strengthens this scenario.

\subsection{\xray~observations}
\label{ss:counterparts_xray}

VHE PWNe may also be detected in the \xray~domain, as exemplified by Vela~X, for which the VHE emission 
matches quite well the extended structure seen by \rosat~and \asca~\citep{c:velax,c:horns07}. Such an
extended \xray~emission has also been measured recently in the direction of HESS~J1825-137 \citep{c:u08} 
and HESS~J1809-193 \citep{c:kp07}. \xmm~and \chan~observations towards \hessj~have been performed and 
analyzed by \citet{c:esposito07} and \citet{c:zavlin07}, but no clear evidence of a PWN has been found yet, 
with the exception of a faint tail-like emission reported by the latter group.

\subsubsection{\rosat}
\label{ss:counterparts_rosat}

\begin{center}
\begin{figure}[!htb]
\centering

  \includegraphics[width=0.48\textwidth]{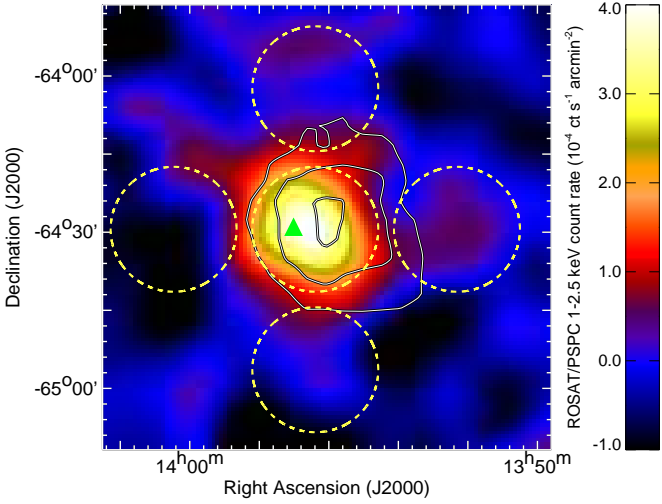}

  \caption{\rosat/PSPC image of the vignetting-corrected and background-subtracted count 
           rate in the 1-2.5\un{keV} band. The image has been smoothed with a Gaussian of
           width $\sigma$ = 5\m. The white contours denote the levels of the \hess~correlated 
           significance, as in Figure \ref{f:ima1}. The position of the pulsar \psr~is marked 
           with a green triangle. The yellow dashed circles show the source and background 
           regions used to estimate the significance of the extended \xray~emission.
	  }		
  \label{f:ima4}

\end{figure}
\end{center}

\vspace{-0.5cm}

\rosat/PSPC archival data\footnote{\texttt{\tiny http://www.xray.mpe.mpg.de/cgi-bin/rosat/data-browser}},
with an average angular resolution of 45\s (FWHM, according to the Calibration Memo CAL/ROS/93-015),
have been inspected in order to search for diffuse \xray~emission towards \hessj. Figure \ref{f:ima4} 
shows the \rosat/PSPC image (smoothed with a Gaussian with $\sigma$ = 5\m) of the vignetting-corrected and
background-subtracted count rate in the 1-2.5\un{keV} band, centered on \hessj. The observed extended and faint 
emission cannot be explained by point-like sources since only two very faint sources, flagged as true detections, 
were found within 0.2\d~from the centroid of the extended emission, in the ROSAT All-Sky Survey Source Catalogs 
\citep{c:voges99,c:voges00}: 1RXS~J135615.9-642757 is marked with a very low reliability, while 1RXS~J135605.5-642902 
has a faint count rate of (4.81 $\pm$ 1.33) $\times$ 10$^{-2}$ s$^{-1}$. In order to calculate the flux and 
significance of this extended \xray~emission, a source region and four background regions were defined, as 
depicted by the yellow dashed circles of 0.2\d~in radius in Figure \ref{f:ima4}. 

After summing counts within the source (C$_{src}$) and background (C$_{bkg}$) regions, the excess count rate was 
calculated as C$_{src}$ - $\alpha$ $\times$ C$_{bkg}$, where $\alpha$ represents the ratio of the exposure times in the 
source and background regions. Since the \rosat/PSPC exposure map of the region centered on \hessj~is quite uniform 
(with a mean value of $\sim$ 0.45\un{ks} and a dispersion of about 3\%), $\alpha$ is simply 1/4. The 
associated significance was computed according to the prescription of \citet{c:lima83}. The excess count 
rate of the \xray~diffuse emission is found to be 0.142 s$^{-1}$ at the 6 $\sigma$ confidence level.
Varying the sizes of the source region between 0.15 and 0.25\d~does not change the results significantly. 
A 2D symmetrical Gaussian fit was applied to the \rosat/PSPC image to quantify the significance of the observed 
source extent. The best-fit Gaussian width is found to be 0.14\d~$\pm$ 0.04\d~(90\% confidence level), much larger than 
the PSPC angular resolution. This extended emission does not appear in the images at lower energies (\ie~in the 0.11-0.4 and 
0.4-1\un{keV} bands), most likely because of the relatively large hydrogen column density. Therefore, its 
flux has been calculated through the HEASARC software WebPIMMS\footnote{\texttt{\tiny http://heasarc.gsfc.nasa.gov/Tools/w3pimms.html}}. 
With $\textit{N}_{\mathrm{H}}$ = (2--7) $\times$ 10$^{21}$ cm$^{-2}$, as measured by \citet{c:esposito07} towards 
\psr, and assuming a power-law with spectral index between 2 and 2.4 (as measured by \hess, see also section 
\ref{s:discu}), the \rosat/PSPC count rate of $\sim$ 0.142 s$^{-1}$ translates into an unabsorbed 1-2.5\un{keV} 
flux of 3.6$^{+1.6}_{-0.9}$ $\times$ 10$^{-12}$ erg cm$^{-2}$ s$^{-1}$.

\subsubsection{\xmm}
\label{ss:counterparts_xmm}

Due to the low level of statistics in the ROSAT data, no spectral information on the nebula could be derived. 
In order to constrain the spectral properties of the extended \xray~emission seen by PSPC, a 85\un{ks} 
archival \xmm~observation (PI : G. Pavlov, ObsId: 0603280101), centered on \psr, was analyzed using SAS v9.0.0. 
To remove the proton flare contamination during the observation, a histogram of the 10--12\un{keV} count rates 
of the two MOS cameras was built. A Gaussian fit was then performed on this 10--12\un{keV} count rate 
distribution in order to remove time intervals where the count rates were beyond 3 $\sigma$ from the mean 
value \citep{c:pa02}. The remaining exposure time after flare screening is 55\un{ks}. Note that, for this observation, 
the PN camera was operated in timing mode and is therefore not used in the analysis.
The instrumental background was derived from the compilation of filter wheel closed observations 
\footnote{\texttt{\tiny http://xmm2.esac.esa.int/external/xmm\_sw\_cal/background/
filter\_closed/}}, in the same 
detector area as the source, and renormalized in the 10--12\un{keV} band over the whole field of view. Events from 
the list of the 42 point sources (\psr~included), automatically detected in the observation by the 
\xmm~Survey Science Center (for a total 0.5-6\un{keV} flux of 1.4 $\times$ 10$^{-12}$ erg cm$^{-2}$ s$^{-1}$), 
were removed in order to focus on the properties of the diffuse \xray~emission.

\begin{center}
\begin{figure}[!htb]
\centering

  \includegraphics[width=0.49\textwidth]{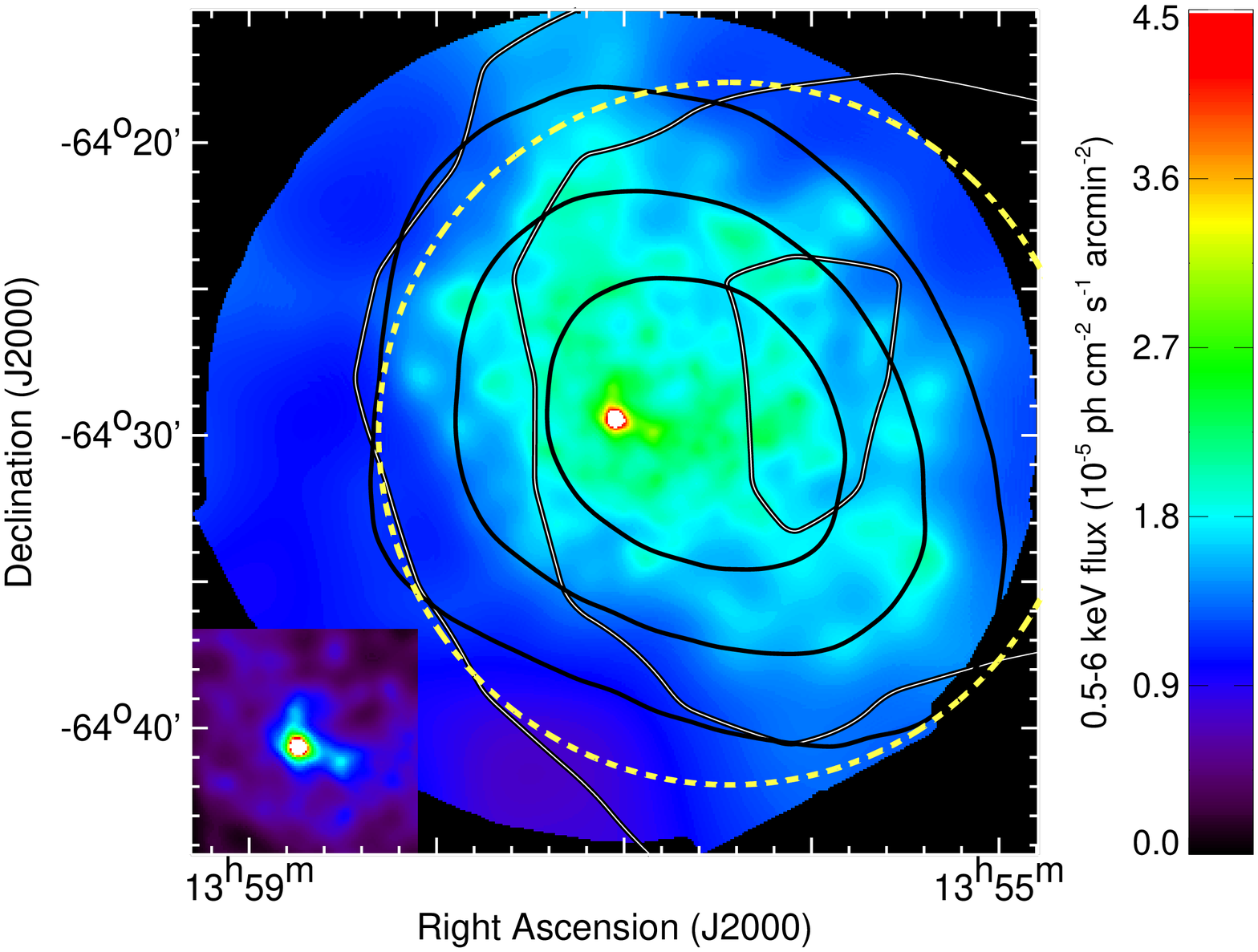}

  \caption{\xmm/MOS image of the 0.5-6\un{keV} flux, adaptively smoothed to a signal-to-noise ratio of 15 (using
           {\it asmooth} from the SAS software) with all the detected point sources, except \psr, removed. Small-scale
           structures close to \psr~are better revealed in the inset image with a colour scale ranging from 0 to 
           8 $\times$ 10$^{-5}$ ph cm$^{-2}$ s$^{-1}$ arcmin$^{-2}$. The solid white contours denote the 
           levels of the \hess~correlated significance, while the black contours correspond to the \rosat/PSPC~smoothed 
           excess count rates. The yellow dashed circle shows the source region used for spectral extraction, as in Figure \ref{f:ima4}.
	  }		
  \label{f:ima5}

\end{figure}
\end{center}

\vspace{-0.5cm}

The vignetting-corrected and background-subtracted combined image of the MOS~1 and MOS~2 cameras in the 0.5--6\un{keV}
band is shown in Figure \ref{f:ima5}. An extended emission is clearly seen around the position of \psr, matching the
extended emission seen with \rosat/PSPC. Small-scale structures are also seen close to \psr, and are discussed in
more details, using \chan~observations, in the companion paper of \citet{c:lemoine11}.

The yellow dashed circle shown in Figure \ref{f:ima5} represents the region of spectral extraction for the \xray~nebula, 
which corresponds to the \rosat/PSPC source extraction region shown in Figure \ref{f:ima4}. The astrophysical background was
extracted over the rest of the field-of-view. The latter spectrum is well described by an absorbed 
APEC\footnote{Astrophysical plasma emission code, see http://hea-www.harvard.edu/APEC} model ($kT$~=~0.21\un{keV},
\textit{norm} = 4.4 $\times$ 10$^{-3}$ cm$^{-5}$) + power-law ($\Gamma$~=~1.95, $N_{0}$ = 7.9 $\times$ 10$^{-4}$ 
cm$^{-2}$ s$^{-1}$ keV$^{-1}$ at 1 keV) model, with $\chi^2$ = 711 for 624 dof. The hydrogen column density was fixed for both 
components to the best-fit value derived by \citet{c:lemoine11} from \chan~observations of \psr~$\textit{N}_{\mathrm{H}}$ = 
($3.9 \pm 0.4$) $\times$ 10$^{21}$ cm$^{-2}$. When let free in our model, the best-fit column density  is $\textit{N}_{\mathrm{H}}$ 
= ($2.9 \pm 0.9$) $\times$ 10$^{21}$ cm$^{-2}$ (errors are quoted at the 90\% confidence level), similar to the value derived 
on the pulsar.

\begin{center}
\begin{figure}[!htb]
\centering

 \includegraphics[width=0.48\textwidth]{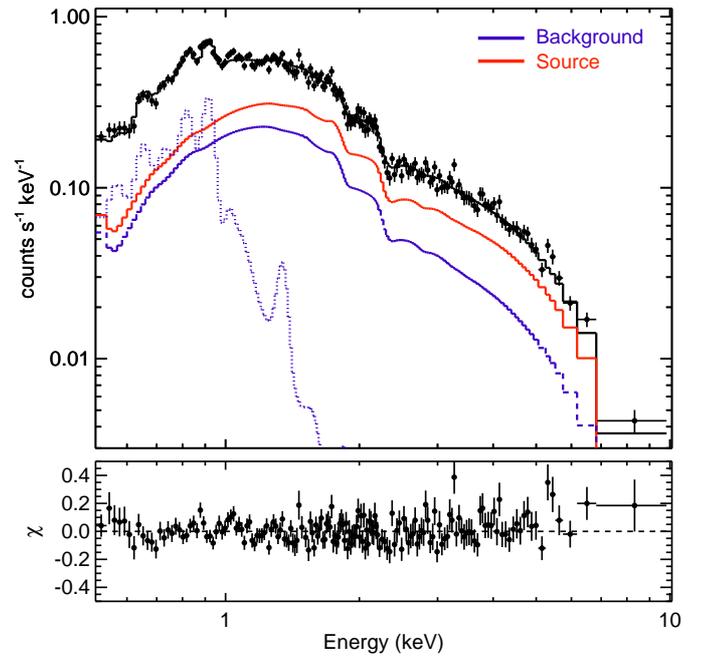}

  \caption{\xmm/MOS 0.5--10\un{keV} spectrum of the total emission (in black) in the source region shown
           in Figure \ref{f:ima5}. The background spectrum (in blue) is best described by an absorbed 
           power-law + APEC model, while the source spectrum (in red) is fitted with an absorbed 
           power-law. A rebinning was applied in order to reach 10 $\sigma$/bin. Residuals are shown
           in the bottom panel.}		
  \label{f:ima6}

\end{figure}
\end{center}

These two best-fit models have been renormalized to the area of the source region and were used as fixed components in the 
fit of the spectrum of the source region (see Figure \ref{f:ima6}). The spectrum of the \xray~nebula (in red) is
described by an absorbed power-law with the column density fixed to the value derived on the pulsar, $\Gamma$ = $1.82 \pm 0.04$ and 
$N_{0}$ = ($2.65 \pm 0.07$) $\times$ 10$^{-3}$ cm$^{-2}$ s$^{-1}$ keV$^{-1}$ at 1 keV, with $\chi^2$ = 683 for 631 dof. The 
unabsorbed 1-2.5\un{keV} flux then amounts to ($4.25 \pm 0.20$) $\times$ 10$^{-12}$ erg cm$^{-2}$ s$^{-1}$, which is in agreement 
with the \rosat/PSPC flux estimated in section \ref{ss:counterparts_rosat}, within the respective uncertainties.

In order to estimate the level of systematics arising from such a large \xray~emission ($\sim$ 0.2\d~in radius, of 
the same order as the \xmm~field of view), several source positions around \psr~have been tested. The resulting
systematic uncertainties on the nebula's best-fit parameters are 0.1 on the photon index and 10\% on the normalization.

\vspace{-0.5cm}

\subsection{\fermi-LAT~observations}
\label{ss:counterparts_he}

After nearly three years of observation, the LAT instrument onboard \fermi~has been successful in identifying the 
HE counterpart of several PWNe powered by energetic PSRs, all of them being also detected in the VHE domain 
\citep[see][and references therein]{c:slane10,c:fermi_pwn}. The high spin-down flux at Earth of \psr~($\simeq$ 5.4 
$\times$ 10$^{35}$ d$^{-2}_{2.4}$ erg s$^{-1}$ kpc$^{-2}$) is in the range of the previously LAT-detected PWNe 
\citep{c:fermi_pwn}. \citet{c:lemoine11} have recently discovered $\gamma$-ray pulsations from \psr~using more than 
two years of \fermi-LAT data, together with radio rotational ephemerides obtained with the Parkes radiotelescope. 
Upper limits on the emission from the associated PWN \hessj~in the 100\un{MeV} -- 100\un{GeV} range have also been 
reported, by assuming a Gaussian shape with $\sigma = $0.2\d, as measured with \hess~These 95\% CL (confidence
level) upper limits, shown in Figure \ref{f:ima8}, were derived using off-pulse data below 4.5\un{GeV}, and from the 
whole dataset above 4.5\un{GeV}, where no pulsed emission is detected.

\section{Discussion}
\label{s:discu}

It was shown in the previous section that an extended radio structure, originally catalogued as a SNR 
candidate, and an extended \xray~emission lie close to the newly discovered VHE source \hessj. Such
radio-\xray-VHE association is usually the observational feature of shell-type and plerionic SNRs. 
Regarding the morphology alone, the current VHE data do not allow us to firmly distinguish between 
these two scenarios. Despite the high confidence level of the \hess~detection, the VHE emission could 
only be characterized as a 0.2\d~symmetric Gaussian. The extended emission seen at 843\un{MHz} appears 
as a faint blob of $\sim$ 0.18\d~in radius with an average flux density of only twice as high as the rms 
noise, while that seen at 2.4\un{GHz}, catalogued as \snc, lies within a complex diffuse emission, 
potentially associated with another SNR candidate. Only the radio emission observed at 4.85\un{GHz} is
safely detected. The \rosat/PSPC extended emission in the 1-2.5\un{keV} band roughly exhibits the same
size as those in the radio and VHE domains, but given its significance of 6 $\sigma$, it could not be
well characterized neither morphologically nor spectrally. However, thanks to a deep \xmm~observation 
towards \psr, this faint and extended \xray~emission has been clearly detected. Therefore, the presence 
of a nearby energetic and young pulsar \psr, the spectral index of the extended structure measured in radio 
and the \xray-VHE association all point towards the scenario of a PWN associated with the past evolution 
of \psr. In the following, it is then assumed that the radio and \xray~sources are really the counterparts 
of \hessj, and the reliability of such a scenario based on qualitative arguments will be discussed, 
keeping in mind that further multi-wavelength observations are needed in order to trigger more 
detailed calculations.

\begin{center}
\begin{figure*}[!htb]
\centering

  \includegraphics[width=0.80\textwidth]{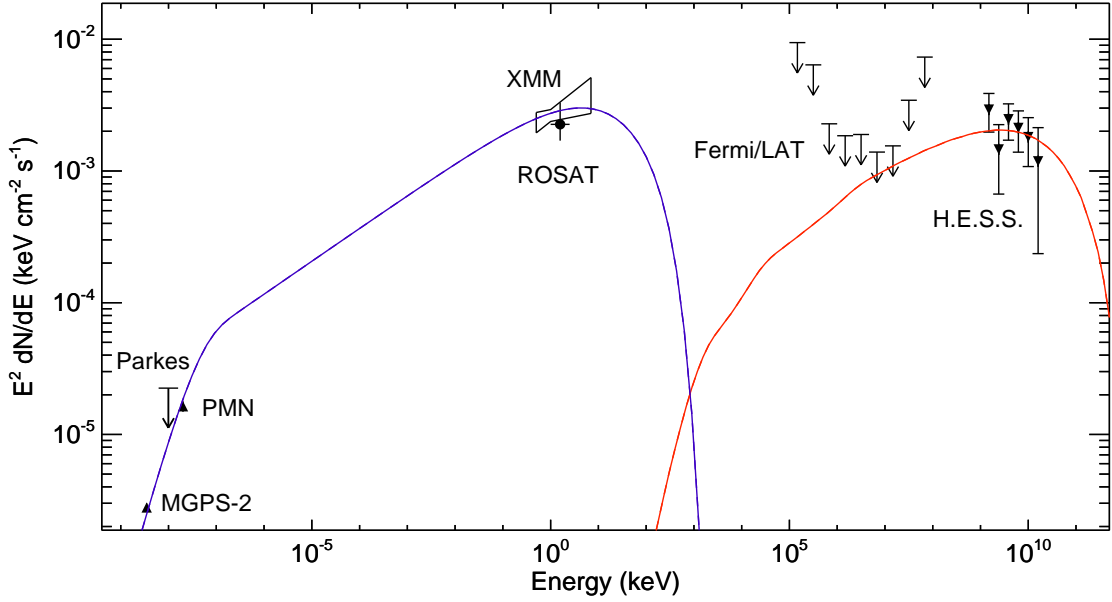}

  \caption{Spectral Energy Distribution of \hessj. The blue and red lines represent the SYN and IC emissions respectively,
           derived from a parent electron spectrum in the form of a power-law with an exponential cutoff (see text). 
           The multi-wavelength data are presented in sections \ref{s:obsres} and \ref{s:counterparts}. The magnetic 
           field strength is set to 3.5 $\mu$G.}		
  \label{f:ima8}

\end{figure*}
\end{center}

\vspace{-0.5cm}

In a one-zone leptonic model, a distribution of accelerated electrons cools radiatively by means of 
synchrotron (SYN) and inverse-Compton (IC) channels in a single volume. The IC scattering is assumed to take 
place in the Thomson regime. Within this emission volume, a constant and entangled magnetic field and a 
homogeneous seed photon distribution from the Cosmic Microwave Background Radiation (hereafter, CMB) only are 
considered. For a power-law distribution of accelerated electrons $K \gamma^{-p}$, the general equation relating 
the SYN energy flux (F$_{SYN}$) produced by electrons with Lorentz factors between $\gamma_{1,SYN}$ and 
$\gamma_{2,SYN}$ and the IC energy flux (F$_{IC}$) radiated between $\gamma_{1,IC}$ and $\gamma_{2,IC}$ 
can be expressed as in \citet{c:rcw86}:

\begin{eqnarray}
\frac{F_{SYN}}{F_{IC}} & = & \frac{U_B}{U_{CMB}} \left(\frac{\gamma_{2,SYN}^{3-p} - \gamma_{1,SYN}^{3-p}}{\gamma_{2,IC}^{3-p} - \gamma_{1,IC}^{3-p}}\right)
\label{e:eq1}
\end{eqnarray}

\noindent where U$_B$ ($\sim$ 2.5 B$^{2}_{-5}$ eV cm$^{-3}$, where B$_{-5}$ is the magnetic field value in units 
of 10 $\mu$G) and U$_{CMB}$ (= 0.26 eV cm$^{-3}$) are the energy densities of the magnetic field and of the CMB photon 
field, respectively. It is further assumed that electrons emit synchrotron and inverse-Compton photons at the
following characteristic energies \citep[the so-called $\delta$-functional approximation, see][]{c:aharonian97}:

\begin{eqnarray*}
\gamma_{SYN} & \simeq & 1.4\times{10^{8}} E_{SYN,keV}^{1/2} B_{-5}^{-1/2} \\
\gamma_{IC}  & \simeq & 3.6\times{10^{7}} E_{IC,TeV}^{1/2} ,              
\end{eqnarray*}

\noindent where E$_{SYN,keV}$ and E$_{IC,TeV}$ are the SYN and IC photon energies in units of keV and TeV, respectively. 
Substituting the above relations into equation \ref{e:eq1} leads to the general expression of the magnetic
field:

\begin{eqnarray}
B_{-5} & = & G(\Gamma) \times{\left(\frac{F_{SYN}}{F_{IC}} \times \frac{(E_{2,IC,TeV}^{2-\Gamma} - E_{1,IC,TeV}^{2-\Gamma})}{(E_{2,SYN,keV}^{2-\Gamma} - E_{1,SYN,keV}^{2-\Gamma})}\right)^{1/\Gamma}}
\label{e:eq2}
\end{eqnarray}

\noindent where G($\Gamma$) $\simeq$ (0.1 $\times$ 15$^{\Gamma-2}$)$^{1/\Gamma}$, and $\Gamma = (p+1)/2$ 
is the common photon index (dN/dE $\propto$ E$^{-\Gamma}$) of SYN and IC spectra. In case the SYN and IC fluxes both 
originate from the same population of electrons, and for $\Gamma$ = 2 (\ie~p~=~3), equation \ref{e:eq2} reduces to 
the standard formula: F$_{SYN}$/F$_{IC}$ $\simeq$ 10 B$^{2}_{-5}$ \citep{c:aharonian97}.

Within this simple leptonic model, the VHE (\hess) and \xray~(\rosat/PSPC and \xmm) measurements then permit an 
exploration of the acceptable range of B-field values in \hessj. Since \rosat/PSPC does not provide any spectral index,
$\Gamma$ is simply assumed to be that measured by \hess~$\Gamma_{VHE}$ = 2.2 $\pm$ 0.2. The \xray~flux measured in
the 1-2.5\un{keV} energy range is provided in section \ref{ss:counterparts_rosat}. As for the \xmm~results, the 
photon index $\Gamma_X$ = 1.82 $\pm$ 0.11\footnote{Statistical and systematical errors on the spectral index and
normalization, provided in section \ref{ss:counterparts_xmm}, have been summed quadratically.} is marginally consistent 
with $\Gamma_{VHE}$ at the 2 $\sigma$ confidence level, so that $\Gamma$ is supposed to lie between 1.8 and 2. 
Accounting for all the uncertainties in the VHE and \xray~measurements, equation \ref{e:eq2} leads to a B-field 
of 4.5 (2.5-8.5) $\mu$G from \rosat/PSPC and 3.5 (2.5-4.5) $\mu$G from \xmm.

The extended structure seen in the \xray~and VHE $\gamma$-ray domains would then represent the ancient PWN
from the early phases of \psr, where electrons which had been accelerated in the past simply age nowadays 
through radiative losses. Thus, a maximal energy is obtained by equating the PSR age (= 2 $\tau_{c}$ / (n$-$1), 
with n the braking index) to the total radiation lifetime \citep[see equation 6 in][]{c:dd09}. The associated 
maximal energies of IC and SYN photons then read:

\begin{eqnarray}
E_{\mathrm{IC, max}} & \sim & \left(\frac{\tau_{c}}{10^{5}~\mathrm{yr}}\right)^{-2} \times{\left(\frac{n-1}{2 + 7.2 (B/5\mu G)^{2}}\right)^{2}}~\mathrm{TeV}
\label{e:eq3}
\end{eqnarray}
\begin{eqnarray}
E_{\mathrm{SYN, max}} & \sim & 0.033~\left(\frac{E^{\gamma}_{\mathrm{IC, max}}}{\mathrm{1~TeV}}\right) \times{\left(\frac{B}{5\mu G}\right)}~\mathrm{keV}
\label{e:eq4}
\end{eqnarray}

For a B-field in the nebula of 3.5 $\mu$G and a PSR age of 7.3\un{kyr} (with n = 3), the IC and SYN photon maximal energies 
would then lie at $\sim$ 17\un{TeV} and $\sim$ 0.5\un{keV}, respectively. The latter value seems too low to account for the 
\xray~spectrum as measured with \xmm, well fitted with a power-law up to $\sim$ 7\un{keV}. However, it should be noted that 
the relevant timescale $\tau$ in such E$_{max}$ estimates is rather the time elapsed since the end of the PWN crushing 
phase, which typically ends a few kyrs after the pulsar birth \citep{c:blondin01,c:vds04}, \ie at the time the ancient PWN 
such as \hessj~is formed. Therefore, for a B-field of 3.5 $\mu$G, equations \ref{e:eq3} and \ref{e:eq4} give 
E$_{\mathrm{IC, max}}$ $\sim$ 200 $(\tau/2.5~\mathrm{kyr})^{-2}$ TeV and E$_{\mathrm{SYN, max}}$ $\sim$ 5 
$(\tau/2.5~\mathrm{kyr})^{-2}$~keV, compatible with the \xmm~and \hess~measurements. Such a maximal energy in the VHE 
domain, at 13.8 $\pm$ 2.3\un{TeV}, has been detected in Vela~X \citep{c:velax}. This measurement led \citet{c:dd09} to 
constrain the magnetic field in the nebula to be as low as $\sim$ 3 $\mu$G, comparable to what is found here for \hessj. 
The VHE spectrum of the latter is well fitted with a power-law between 1 and 20\un{TeV}, and the 95~\% CL lower limit on 
a cutoff energy of 3.5\un{TeV} derived in section \ref{s:obsres} is compatible with these estimates.

Within this one-zone leptonic framework, the exact expressions of SYN and IC emissions from a parent electron spectrum 
have been calculated \citep{c:bg70} and qualitatively fitted to the multi-wavelength data for a B-field of 3.5 $\mu$G, as 
shown in Figure \ref{f:ima8}. Besides the CMB, the Galactic interstellar radiation field at the location of \hessj, used to 
calculate the IC emission, was derived from the latest estimates of 
\citet{c:porter05}\footnote{\texttt{\tiny http://galprop.stanford.edu/resources.php?option=data}}. It comprises the Galactic 
infrared (from dust, at T $\sim$ 35 K and 350 K) and optical (from stars, at T $\sim$ 4600 K) emission, with energy 
densities of 0.66 and 0.94 eV cm$^{-3}$, respectively. In addition to the radio, \xray~and VHE \gammaray~data, the \fermi-LAT 
upper limits discussed in section \ref{ss:counterparts_he} \citep{c:lemoine11}, are also shown. The parent electron spectrum 
is in the form dN$_e$/dE$_e$ $\propto$ E$_e$$^{-p}$ exp($-$E$_e$/E$_{\rm cut}$) for E$_e$ $\in$ $[E_{\rm min},E_{\rm max}]$, 
with spectral index p = 2.5, E$_{\rm min}$ = 13\un{GeV} (and E$_{\rm max}$ set to 1 PeV), E$_{\rm cut}$ = 350 TeV, and a total 
lepton energy  E$_{\rm tot}$~=~5~$\times$~10$^{47}$~d$^{2}_{2.4}$ erg. Note that a low energy cutoff of 13\un{GeV} in the 
electron spectrum is in the range of minimum particle energies considered in several works \citep{c:kc84,c:fermi_pwn}. 
The total particle energy must be compared to the rotational kinetic energy of \psr~since birth, 
E$_{sd}$ = ($\Omega_0^2$ $-$ $\Omega_t^2$)$I$/2, where the spin periods $P_i$ = 2$\pi$/$\Omega_i$ (the indices $0$ and $t$ 
denote the initial and current values, respectively) and $I$ is the moment of inertia of the neutron star (taken to be 1.4 
$\times$ 10$^{45}$ g cm$^2$). With $P_t$ = 166 ms, a lower limit on the rotational kinetic energy 
E$_{sd}$ $\gtrsim$ 5 $\times$ 10$^{48}$ erg is derived for any intial period $P_0$ $\lesssim$ 70 ms. Therefore, this spin-down 
energy implies a reasonable conversion efficiency $\eta = $ E$_{\rm tot}$/E$_{sd}$ $\lesssim$ 0.1 d$^{2}_{2.4}$.

It should be noted that these calculations implicitely assume that the observed radio, \xray~and VHE morphologies
indeed probe the same emissive region, whereas in Figures \ref{f:ima4} and \ref{f:ima5}, there seems to be a 
slight offset between the \rosat/\xmm~and \hess~extended emissions, \xrays~coming from a more compact region closer 
to \psr, \ie where the magnetic field could be higher. Only through detailed calculations which take into account the 
spatial and temporal evolution of both particles and magnetic field in such PWN system \citep[see {\it e.g.}][]{c:gelfand09}, 
could these key parameters be efficiently constrained.

\section{Conclusion}
\label{s:conclu}

A new VHE source, namely \hessj, has been discovered during the continuation of the \hess~Galactic Plane
Survey. This extended source lies close to \psr, a recently discovered young ($\tau_c$ = 7.3\un{kyr}), 
nearby (d = 2.4\un{kpc}) and energetic pulsar (\edot~= 3.1 $\times$ 10$^{36}$ erg s$^{-1}$). Archival radio and 
\xray~data have revealed an extended structure, though faint, coincident with the VHE emission. Given the 
faintness of the emission seen in these energy domains, the poor information on its morphology prevents one from 
distinguishing between shell-type and plerionic SNR as the origin of \hessj. However, (1) the presence of 
an energetic pulsar, (2) the centre-filled morphology with a flat spectral index measured in radio, and (3) 
the existence of a diffuse \xray~emission point towards the scenario of an evolved and offset VHE PWN.
 
Besides comparable ages, \hessj~shares several similarities with the Vela~PWN (whose pulsar PSR B0833$-$45 has a 
$\tau_c$ = 11.3\un{kyr} and lies at a distance of 290\un{pc}), regarding both the pulsar (very low L$_{X}$/\edot~efficiencies 
and presence of thermal \xray~emission) and the PWN (similar ratios of compact/X-ray to diffuse/radio PWN sizes) 
properties. The 2\s~tail-like \xray~emission close to \psr~\citep{c:zavlin07} could be the pulsar jet, 
which is observed in the Vela pulsar at the sub-arcmin scales. The radio counterpart of \hessj, $\sim$ 0.35\d~in 
size, would thus correspond to the bright 2\d~$\times$ 3\d~radio structure (the so-called Vela~X), extending  
southwest of the Vela pulsar. In terms of the broad-band non-thermal emission, the so-called Vela cocoon, a 
45\m-long collimated \xray~filament extending southward of PSR B0833$-$45, has been detected in VHE $\gamma$-rays 
\citep{c:velax}, while the whole Vela~X radio structure was recently detected in the GeV domain with
\agile~\citep{c:agile10} and \fermi-LAT \citep{c:velaX10}. These measurements seem to point towards the scenario
of two distinct lepton spectral components, first suggested by \cite{c:dejager07}, and later studied by 
\cite{c:dejager08} and \cite{c:lamassa08}. However, follow-up \hess~observations have recently allowed for 
the detection of significant VHE emission beyond the cocoon, up to 1.2\d~from the VHE barycentre, with a spectrum
similar to that of the cocoon \citep{c:dubois09}. In the case of \hessj, at the estimated distance of \psr, the 
putative ``cocoon-like'' structure would appear as nearly point-like ($\sim$ 0.09\d~long) for \hess~However, such 
structure cannot be the dominant VHE component as \hessj~is well fitted simply with a Gaussian width of 0.2\d~(see
section \ref{s:obsres}). \hessj~would then represent the VHE counterpart to the ``Vela X-like'' radio structure 
(see section \ref{ss:counterparts_radio}), detected also in \xrays~with \rosat/PSPC and \xmm~(see section 
\ref{ss:counterparts_xray}). Thus, given the observational limitations of the current radio and VHE 
\gammaray~data, the broad-band spectrum of \hessj~can be explained by a single lepton population with reasonable 
parameters, as presented in section \ref{s:discu}.

From qualitative arguments, it was shown that the magnetic field within \hessj~must be quite low ($\sim$ 3--4 $\mu$G), 
of the same order as what has been implied in other PWNe \citep[\eg Vela~PWN, see][]{c:dejager08}, under 
similar assumptions as investigated here. Follow-up observations in radio will help to constrain the pulsar 
braking index and proper motion, and confirm the nature of the extended emission. The combined GeV-TeV data from 
the continuation of Fermi observations and future VHE observations with the upcoming \hess~phase II will better
constrain the spectral shape of the IC emission, in order to probe the magnetic field strength and the energetics
of \hessj.


\begin{acknowledgements}
We would like to thank Frank Haberl for his help and precious advice on the \rosat~archival images,
and Tara Murphy for helpful discussion about the MPGS-2 radio data. The support of the Namibian 
authorities and of the University of Namibia in facilitating the construction and operation of H.E.S.S. 
is gratefully acknowledged, as is the support by the German Ministry for Education and Research (BMBF), 
the Max Planck Society, the French Ministry for Research, the CNRS-IN2P3 and the Astroparticle 
Interdisciplinary Programme of the CNRS, the U.K. Science and Technology Facilities Council (STFC), the 
IPNP of the Charles University, the Polish Ministry of Science and Higher Education, the South African 
Department of Science and Technology and National Research Foundation, and by the University of Namibia. 
We appreciate the excellent work of the technical support staff in Berlin, Durham, Hamburg, Heidelberg, 
Palaiseau, Paris, Saclay, and in Namibia in the construction and operation of the equipment.
\end{acknowledgements}


\bibliographystyle{aa}

\begin{thebibliography}{}

\bibitem[Abdo \etal, 2010b]{c:velaX10} Abdo, A.~A., \etal (\fermi~collaboration) 2010b, \apj, 713, 146
\bibitem[Aharonian \etal, 1997]{c:aharonian97} Aharonian, F., Atoyan, A.~M., \& Kifune, T. 1997, \mnras, 291, 162
\bibitem[Aharonian \etal, 2005a]{c:g09}     Aharonian, F., \etal (\hess~collaboration) 2005a, \aap, 432, L25
\bibitem[Aharonian \etal, 2005b]{c:msh1552} Aharonian, F., \etal (\hess~collaboration) 2005b, \aap, 435, L17
\bibitem[Aharonian \etal, 2006b]{c:survey1} Aharonian, F., \etal (\hess~collaboration) 2006b, \apj, 636, 777
\bibitem[Aharonian \etal, 2006c]{c:velax}   Aharonian, F., \etal (\hess~collaboration) 2006c, \aap, 448, L43
\bibitem[Aharonian \etal, 2006d]{c:crab}    Aharonian, F., \etal (\hess~collaboration) 2006d, \aap, 457, 899
\bibitem[Aharonian \etal, 2006e]{c:j1825}   Aharonian, F., \etal (\hess~collaboration) 2006e, \aap, 460, 365
\bibitem[Aharonian \etal, 2007b]{c:j1809}  Aharonian, F., \etal (\hess~collaboration) 2007b, \aap, 472, 489
\bibitem[Aharonian \etal, 2008a]{c:darks} Aharonian, F., \etal (\hess~collaboration) 2008a, \aap, 477, 353
\bibitem[Aharonian \etal, 2008b]{c:j1912} Aharonian, F., \etal (\hess~collaboration) 2008b, \aap, 484, 435
\bibitem[Aharonian \etal, 2009]{c:rcw86} Aharonian, F., \etal (\hess~collaboration) 2009, \apj, 692, 1500
\bibitem[Ackermann \etal, 2011]{c:fermi_pwn} Ackermann, M., \etal 2011, \apj, 726, 35
\bibitem[Amato \etal, 2003]{c:amato03} Amato, E., Guetta, D., \& Blasi, P. 2003, \aap, 402, 827
\bibitem[Atwood \etal, 2009]{c:fermi09} Atwood, W.~B., \etal (\fermi~collaboration) 2009, \apj, 697, 1071
\bibitem[Bednarek \& Bartosik, 2003]{c:bb03} Bednarek, W., \& Bartosik, M. 2003, \aap, 405, 689
\bibitem[Berge \etal, 2007]{c:berge07} Berge, D., \etal 2007, \aap, 466, 1219
\bibitem[Blondin \etal, 2001]{c:blondin01} Blondin, J.~M., Chevalier, R.~A., \& Frierson, D.~M. 2001, \apj, 563, 806
\bibitem[Blumenthal \& Gould, 1970]{c:bg70} Blumenthal, G.~R., \& Gould, R.~J. 1970, Rev. Mod. Phys., 42, 237
\bibitem[Bucciantini, 2008]{c:bucciantini08} Bucciantini, N. 2008, AIP Conference Proceedings, Vol. 983, pp. 186-194
\bibitem[Camilo \etal, 2004]{c:camilo04} Camilo, F., \etal 2004, \apj, 611, L25
\bibitem[Carter \& Read, 2007]{c:cr07} Carter, J.~A., \& Read, A.~M. 2007, \aap, 464, 1155
\bibitem[Chaves \etal, 2008]{c:survey2} Chaves, R.~C.~G., \etal (\hess~collaboration) 2008, AIP Conference Proceedings, Vol. 1085, pp. 219
\bibitem[Condon \etal, 1993]{c:condon93} Condon, J.~J., Griffith, M.~R., \& Wright, A.~E. 1993, \aj, 106, 1095
\bibitem[Daum \etal, 1997]{c:daum97} Daum, A., \etal (HEGRA~collaboration) 1997, Astropart. Phys., 8, 1
\bibitem[de Jager, 2007]{c:dejager07} de Jager, O.~C. 2007, \apj, 658, 1177
\bibitem[de Jager \etal, 2008]{c:dejager08} de Jager, O.~C., Slane, P.~O., \& LaMassa, S.~M. 2008, \apj, 689, L125
\bibitem[de Jager \& Djannati-Ata\"i, 2009]{c:dd09} de Jager, O.~C., \& Djannati-Ata\"i, A. 2009, Neutron Stars and Pulsars, ASSL, Vol. 357, pp. 451
\bibitem[de Jager \etal, 2009]{c:dejager09} de Jager, O.~C., \etal 2009, Proceedings of the 31st ICRC, in press 
\bibitem[Djannati-Ata\"i \etal, 2008]{c:youngs} Djannati-Ata\"i, A., \etal (\hess~collaboration) 2008, Proceedings of the 30th ICRC, 2, 823
\bibitem[Dubois \etal, 2009]{c:dubois09} Dubois, F., \etal (HESS~collaboration) 2009, Proceedings of the 31st ICRC, in press 
\bibitem[Duncan \etal, 1995]{c:duncan95} Duncan, A.~R., \etal 1995, \mnras, 277, 36
\bibitem[Duncan \etal, 1997]{c:duncan97} Duncan, A.~R., \etal 1997, \mnras, 287, 722
\bibitem[Esposito \etal, 2007]{c:esposito07} Esposito, P. \etal 2007, \aap, 467, L45
\bibitem[Gaensler \& Slane, 2006]{c:gaensler06} Gaensler, B.~M., \& Slane, P.~O. 2006, \araa, 44, 17
\bibitem[Gallant \etal, 2008]{c:gallant08} Gallant, Y., \etal 2008, AIP Conference Proceedings, 983, 195
\bibitem[Gelfand \etal, 2009]{c:gelfand09} Gelfand, J.~D., Slane, P.~O., \& Zhang, W. 2009, \apj, 703, 2051
\bibitem[Green, 2009]{c:green09} Green, D.~A. 2009, BASI, Vol.37, No.1, p.45
\bibitem[Griffith \& Wright, 1993]{c:griffith93} Griffith, M.~R., \& Wright, A.~E. 1993, \aj, 105, 1666
\bibitem[Hillas, 1996]{c:hillas96} Hillas, A.~M. 1996, Space Sci. Rev., 75, 17
\bibitem[Hinton \& Hofmann, 2009]{c:hinton09} Hinton, J.~A., \& Hofmann, W. 2009, \araa, 47, 523
\bibitem[Hobbs \etal, 2005]{c:hobbs05} Hobbs, G., \etal 2005, \mnras, 360, 974
\bibitem[Horns \etal, 2006]{c:horns06} Horns, D., \etal 2006, \aap, 451, L51
\bibitem[Horns \etal, 2007]{c:horns07} Horns, D., \etal 2007, Ap\&SS, 309, 189
\bibitem[Kargaltsev \& Pavlov, 2007]{c:kp07} Kargaltsev, O., \& Pavlov, G.~G. 2007, \apj, 670, 655
\bibitem[Kennel \& Coroniti, 1984]{c:kc84} Kennel, C.~F., \& Coroniti, F.~V. 1984, \apj, 283, 710
\bibitem[LaMassa \etal, 2008]{c:lamassa08} LaMassa, S.~M., Slane, P.~O., \& de Jager, O.~C. 2008, \apj, 689, L121
\bibitem[Lemoine-Goumard \etal, 2011]{c:lemoine11} Lemoine-Goumard, M., \etal 2011, \aap~in press, arXiv:1108.0161
\bibitem[Li \& Ma, 1983]{c:lima83} Li, T.-P., \& Ma, Y.-Q. 1983, \apj, 272, 317
\bibitem[Murphy \etal, 2007]{c:murphy07} Murphy, T., \etal 2007, \mnras, 382, 382
\bibitem[Paladini \etal, 2003]{c:paladini03} Paladini, R., \etal 2003, \aap, 397, 213
\bibitem[Pellizzoni \etal, 2010]{c:agile10} Pellizzoni, A., \etal (\agile~collaboration) 2010, Science, 327, 663
\bibitem[Porter \& Strong, 2005]{c:porter05} Porter, T.~A., \& Strong, A.~W. 2005, Proceedings of the 29th ICRC, Vol. 4, p. 77
\bibitem[Pratt \& Arnaud, 2002]{c:pa02} Pratt, G.~W., \& Arnaud, M. 2002, \aap, 394, 375
\bibitem[Russeil, 2003]{c:russeil03} Russeil, D. 2003, \aap, 397, 133
\bibitem[Slane \etal, 2010]{c:slane10} Slane, P., \etal 2010, \apj, 720, 266
\bibitem[Uchiyama \etal, 2008]{c:u08} Uchiyama, H., \etal 2008, \pasj, Vol. 61, S189--S196
\bibitem[van der Hucht, 2001]{c:vdh01} van der Hucht, K.~A. 2001, New Astr. Reviews, 45, 135
\bibitem[van der Swaluw \etal, 2004]{c:vds04} van der Swaluw, E., Downes, T.~P., \& Keegan, R. 2004, \aap, 420, 937
\bibitem[Voges \etal, 1999]{c:voges99} Voges, W., \etal 1999, \aap, 349, 389
\bibitem[Voges \etal, 2000]{c:voges00} Voges, W., \etal 2000, IAUC Circ., 7432, 3
\bibitem[Wright \etal, 1994]{c:wright94} Wright, A.~E., \etal 1994, \apjs, 91, 111
\bibitem[Zavlin, 2007]{c:zavlin07} Zavlin, V.~E. 2007, \apj, 665, L143

\end{thebibliography}


\end{document}